\documentclass[aps,pra,twocolumn,superscriptaddress,amsmath,showpacs,amssymb]{revtex4} 

\usepackage{graphicx}
\usepackage{amsthm,amsmath}

\newcommand{\bm}[1]{\mbox{\boldmath{$#1$}}}
\newcommand{\be}{\begin{equation}}
\newcommand{\ee}{\end{equation}}
\newcommand{\ba}{\begin{eqnarray}}
\newcommand{\ea}{\end{eqnarray}}

\newcommand{\nn}{\nonumber}

\newcommand{\bmx}{\bm{x}}

\newcommand{\ka}{{\kappa}}

\begin{document}

\title{Condition for the Existence of Complex Modes in a Trapped Bose--Einstein Condensate with a Highly Quantized Vortex}

\author{E.~Fukuyama}
\affiliation{Department of Physics, Waseda University,
 Tokyo 169-8555, Japan}
\author{M.~Mine}
\email{mine@aoni.waseda.jp}
\affiliation{Department of Physics, Waseda University,
 Tokyo 169-8555, Japan}
\author{M.~Okumura}
\email{okumura@aoni.waseda.jp}
\affiliation{Department of Applied Physics, Waseda University,
 Tokyo 169-8555, Japan}
\author{T.~Sunaga}
\email{tomoka@fuji.waseda.jp}
\affiliation{Department of Physics, Waseda University,
 Tokyo 169-8555, Japan}
\author{Y.~Yamanaka}
\email{yamanaka@waseda.jp}
\affiliation{Department of Materials Science and Engineering,
Waseda University, Tokyo 169-8555, Japan}

\date{\today}

\begin{abstract}
We consider a trapped Bose--Einstein condensate (BEC) with a highly
quantized vortex. 
For the BEC with a doubly, triply or quadruply quantized vortex, 
the numerical calculations have shown that the Bogoliubov--de Gennes 
equations, which describe the fluctuation of the condensate, have
complex eigenvalues. 
In this paper, we obtain the analytic expression of the condition for
the existence of complex modes, using the method developed by Rossignoli
and Kowalski [R.~Rossignoli and A.~M.~Kowalski, Phys. Rev. A {\bf 72},
032101 (2005)] for the small coupling constant. 
To derive it, we make the two-mode approximation. With the derived
analytic formula, we can identify the quantum number of the complex
modes for each winding number of the vortex. Our result is consistent
with those obtained by the numerical calculation in the case that the
winding number is two, three or four.  
We prove that the complex modes always exist when the condensate has a 
 highly quantized vortex.  
\end{abstract}


\maketitle

\section{Introduction}\label{intro}

The Bose--Einstein condensates (BECs) of neutral atomic gases were
realized in 1995 \cite{Boulder,MIT,Li}, and several kinds of vortices 
inside the BECs have been observed: the singly quantized vortex
\cite{Matthews}, the vortex lattice \cite{Madison,Abo-Shaeer}, and the
doubly quantized vortex \cite{Leanhardt} which was created by the
topological phase engineering \cite{Nakahara}. 
In particular, an interesting phenomenon was observed: a doubly
quantized vortex decayed into two single quantized vortices
spontaneously \cite{Shin}. 

The theoretical investigations on the instability of the highly
quantized vortex of the neutral atomic BEC were made by
several authors~\cite{Pu,Mettenen,Kawaguchi}. 
They solved the Bogoliubov--de Gennes (BdG) equations
\cite{Bogoliubov,deGennes,Fetter} numerically, and found that the equations
 have complex eigenvalues when a condensate has a vortex with the winding
number two~\cite{Pu, Mettenen}, three~\cite{Pu} or four~\cite{Kawaguchi}.  
The complex eigenvalues cause the blowup or damping of the c-number
condensate fluctuations~\cite{Pu,Mettenen,Kawaguchi}. 
This instability of the condensate, caused by the complex eigenvalues of
the BdG equations, is called ``dynamical instability''. 
But it is still unknown whether a highly quantized vortex with an 
arbitrary high winding number always brings complex eigenvalues and
which eigenvalues become complex. 
It should also stressed that the relation between the
``dynamical instability'' in theoretical concept and the observed decay
of the doubly quantized vortex is not elucidated fully and still under
investigation \cite{Gawryluk,Huhtameki}. 

As for the ``complex modes'', there is known another treatment, developed
by Rossignoli and Kowalski \cite{RK}. 
We refer to it as the RK method. 
In this method the quantum Hamiltonian of the quadratic form of
creation and annihilation operators is considered, and the complex modes
appear as a result of diagonalizing it with unusual operators which are 
neither bosonic nor fermionic ones. 
Obviously the meaning of the complex modes in the RK method is different
from one mentioned above \cite{Pu,Mettenen,Kawaguchi}: they are quantum
fluctuations in the former, while they are c-number ones in the latter.  
The relation between the complex modes and the complex eigenvalues of
the BdG equations has not been established.
Further, we point out that the interpretation of the complex
eigenvalues of the BdG equations in the case that they are treated as
quantum fluctuations, as well as of the complex modes of the RK method,
is not simple.
We recently proposed a new approach of quantum field theory, using the
eigenfunctions of the BdG equations including those belonging to the
complex eigenvalues and considering the free quantum Hamiltonian
consistently \cite{Mine}, although the content of this paper is not
related to the approach directly.

In this paper, we derive the conditions for the existence of the complex
modes in a trapped BEC with a highly quantized vortex analytically using
the RK method within the two-mode approximation and small coupling
expansion. 
Using the condition, we can answer the question above: does any highly
quantized vortex bring the complex modes? The answer is yes, and we can
identify partially which mode is the complex one.  
The condition is also useful for checking the results of the numerical
calculation. 

This paper is organized as follows. In Sec.~\ref{sec-Model}, the model
Hamiltonian describing the weak interacting neutral atoms in a harmonic
trap is introduced. The c-number field whose square represents the
distribution of the condensate is subject to the Gross--Pitaevskii (GP),
and the quantum field is expanded in terms of a complete set.
In Sec.~\ref{sec-RK}, we give a brief review of the RK method.
In Sec.~\ref{sec:condition}, first, we explain the two-mode approximation.
Second, the small coupling expansion is introduced. 
Both of them play essential roles in our analysis. 
We derive the condition for the existence of the complex modes
analytically within the two-mode approximation and small coupling
expansion. 
Finally, we prove that the complex modes always exist when the
condensate has a highly quantized vortex. 
Section \ref{summary} is devoted to the summary.

\section{Model Hamiltonian}\label{sec-Model}
We start with the following Hamiltonian to describe the neutral atoms
trapped by a harmonic potential of a cylindrical symmetry, 
\begin{align}
\label{H}
\hat{H} & = \int \!\! d^3 x \Bigl[ \hat{\psi}^{\dag} (x) ( K + V (r,z) - 
\mu ) \hat{\psi} (x) \nn \\
& \quad {} + \frac{g}{2} \hat{\psi}^{\dag 2} (x) \hat{\psi}^2
(x) \Bigr],
\end{align}
where $r=\sqrt{x^2 + y^2}$, $x=(\bmx,t)$ and
\begin{align}
K & = - \frac{1}{2M} \nabla^2 \, , \\ 
V (r,z) & = \frac{1}{2} M ( \omega_{\perp}^2 r^2 + \omega_z^2  z^2 ) 
\end{align}
with the mass of a neutral atom $M$, the chemical potential $\mu$ and
the coupling constant $g$. The bosonic field operator $\hat{\psi}(x)$
obeys the canonical commutation relations, 
\begin{align}
& [\hat{\psi} (x),\ \hat{\psi}^{\dag} (x')]|_{t = t'} = \delta
 (\bm{x}-\bm{x'}), \\   
&  [\hat{\psi} (x),\ \hat{\psi} (x')]|_{t=t'} 
= [\hat{\psi}^{\dag} (x),\ \hat{\psi}^{\dag} (x')]|_{t = t'} = 0 \, . 
\end{align}
Let us divide the field operator $\hat{\psi} (x)$ into a classical field
$\xi (\bm{x})$ and a quantum field $\hat{\varphi} (x)$ as 
\begin{equation}
\label{psi}
\hat{\psi} (x) = \xi (\bm{x}) + \hat{\varphi} (x) \, .  
\end{equation}
Here, the classical field $\xi (\bm{x})$ is the order parameter
representing the condensate with a (highly) quantized vortex, which is
assumed to be time-independent. 
Note that the function $\xi (\bm{x})$ is essentially complex due to the
existence of the quantized vortex. 
At the tree level, $\xi (x)$ satisfies the following GP equation:  
\begin{equation}
\label{GP}
\left[ K + V (r,z) -\mu + g |\xi (\bm{x})|^2 \right] \xi(\bm{x}) = 0 \, .  
\end{equation}
The condensate particle number $N_{\rm c}$ is given by 
\begin{equation}
\int \! d^3 x \, |\xi (\bmx)|^2 = N_{\rm c} \, .  
\end{equation}
Substituting Eq.~(\ref{psi}) into Eq.~(\ref{H}), one can rewrite the
total Hamiltonian $\hat{H}$ as 
\begin{equation}
\hat{H} = \hat{H}_0 + \hat{H}_{\rm int} \, ,  
\end{equation}
where the free and interaction Hamiltonians are chosen, respectively,
as 
\begin{align}
\hat{H}_0 & = \int \! d^3 x \, \Bigl[ \hat{\varphi}^{\dag} (x) \left\{ K
 + V (r,z) - \mu \right\} \hat{\varphi} (x) \nn \\  
& \quad {} + \frac{g}{2} \bigl\{ 4 |\xi (\bmx)|^2 \hat{\varphi}^{\dag}
 (x) \hat{\varphi} (x) \nn \\
& \quad {} + \xi^{\ast 2} (\bmx) \hat{\varphi}^2 (x) + \xi^2 (\bmx)
 \hat{\varphi}^{\dag 2} (x) \bigr\} \Bigr], \\ 
\hat{H}_{\rm int} & = g \int \! d^3x \, \biggl[ \xi (\bmx)
 \hat{\varphi}^{\dag 2} (x) \hat{\varphi}(x) + \xi^{\ast } (\bmx)
 \hat{\varphi}^{\dag} (x) \hat{\varphi}^2 (x) \nn \\ 
& \quad {} + \frac{1}{2} \hat{\varphi}^{\dag 2} (x)
 \hat{\varphi}^2(x) \biggr] \, . 
\end{align}
Assuming that a vortex is created along the $z$-axis and taking account of
an axial symmetry along the vortex line, we introduce a real function
$f(r, z)$, defined as 
\begin{equation}
\label{cond}
\xi (\bmx) = \sqrt{\frac{N_{\rm c}}{2 \pi}} e^{i \kappa \theta} f(r, z)
\, , 
\end{equation}
where the integer $\kappa \ge 1$ is a winding number of the
vortex. The function $f(r, z)$ is normalized as 
\begin{equation}
\label{normf} 
\int \! rdr \, dz \, f^2 (r,z) = 1 \, . 
\end{equation}
Substituting  Eq.~(\ref{cond}) into Eq.~(\ref{GP}), one obtains 
\begin{equation}
\label{GP2}
\left[K_{\kappa} + V(r,z) - \mu + \frac{gN_{\rm c}}{2 \pi} f^2 (r,z)
\right] f(r,z) = 0 \, , 
\end{equation}
where 
\begin{equation}
 K_{\kappa} = - \frac{1}{2M} \left( \frac{\partial^2}{\partial r^2} +
 \frac{1}{r} \frac{\partial}{\partial r} - \frac{\kappa^2}{r^2} +
 \frac{\partial^2}{\partial z^2} \right) \, . 
\end{equation}

Using the bosonic creation and annihilation operators
$\hat{a}_{n,\ell,m}^\dag$ and $\hat{a}_{n,\ell,m}$, we expand the
bosonic field $\hat{\varphi} (x)$, 
\begin{equation}
\hat{\varphi} (x) = \sum_{n,m=0}^{\infty} \sum_{\ell= - \infty}^{\infty} 
\hat{a}_{n,\ell,m} e^{i (\ell + \kappa ) \theta} u_{n,\ell,m} (r,z) \, , 
\end{equation}
where $n\ (= 0, 1, 2, \cdots)$ is the principle quantum number
(representing the radial nodes), $\ell \ (= 0, \pm 1, \pm 2, \cdots)$ is 
the magnetic quantum number, $m \ (= 0, 1, 2, \cdots)$ is the quantum
number along the $z$ axis. The function $u_{n,\ell,m} (r,z)$ is a
solution of the eigenequation with an eigenvalue $\epsilon_{n,\ell,m}$, 
\begin{equation}
\label{u_eigen}
\left[ K_{\kappa+\ell} + V(r,z) \right] u_{n,\ell,m} (r,z) =
\varepsilon_{n,\ell,m} u_{n,\ell,m} (r,z) \, ,   
\end{equation}
and satisfies the following orthogonal and completeness conditions: 
\begin{align}
& 2 \pi \int \! rdr \,dz \, u_{n,\ell,m} (r,z) u_{n',\ell,m'} (r,z) =
 \delta_{n,n'} \delta_{m,m'} \, , \\ 
& \sum_{n,m=0}^{\infty} u_{n,\ell,m} (r,z) u_{n,\ell,m} (r',z') =
 \frac{1}{2 \pi r} \delta (r-r') \delta(z-z') \, . 
\end{align}
The eigenfunctions in Eq.~(\ref{u_eigen}) are expressed with
the Laguerre polynomials $L_n ^k (x)$ and the Hermite polynomials $H_m
(x)$ \cite{LH} as 
\begin{align}
u_{n,\ell,m} (r,z) & = C_{n,\ell,m} e^{-\frac{1}{2} \alpha_{\bot}^2 r^2}
(\alpha_{\bot} r )^{|\ell + \kappa|} L_{n}^{|\ell + \kappa|}
 (\alpha_{\bot} ^2 r^2) \nn \\  
& \quad {} \times e^{- \frac{1}{2} \alpha_z^2 z^2} H_m (\alpha_z z) \, ,
 \label{un} 
\end{align}
where
\begin{equation}
C_{n,\ell,m} = \sqrt{ \frac{\alpha_{\bot}^2 \alpha_z }{\pi^{3/2}}} 
\cdot \sqrt{ \frac{n!}{2^m \ m! \ (|\ell + \kappa| + n)!} } \, , 
\end{equation}
$\alpha_{\bot} = \sqrt{M \omega_{\bot}}$ and $\alpha_z = \sqrt{M
\omega_z}$. 
The eigenvalues are given as 
\begin{equation}
 \varepsilon_{n,\ell,m} = \omega_{\bot} \left( 2n + |\ell + \kappa|+ 1
\right) + \omega_z \left( m + \frac{1}{2} \right) \, . \label{epsn}
\end{equation}
The free Hamiltonian is a sum of the sectors
$\hat{H}^{|\ell|}$, labeled by the absolute value of the magnetic quantum
number $j = |\ell|$:
\begin{equation}
\hat{H}_0 = \sum_{j=0}^{\infty} \hat{H}^j = \hat{H}^{0} +
 \sum_{j'=1}^{\infty} \hat{H}^{j'} \, .  \label{H_sectors}
\end{equation}
Hereafter we use the symbols $j$ as $j=0,1,2,\cdots$, and $j'$ as
$j'=1',2',\cdots$, to make it explicit whether the special sector with $j=0$
is included or not in the arguments.

The sectors $\hat{H}^{0}$ and $\hat{H}^{j'}$ ($j' \ge 1$) are explicitly 
given as 
\begin{align}
\hat{H}^{0} & = \sum_{n,n',m,m'=0}^{\infty} \biggl[
 \hat{a}_{n,0,m}^{\dag} \hat{a}_{n',0,m'} A (n,0,m;n',0,m') \nn \\
& \quad {} + \frac{1}{2} \left\{ \hat{a}_{n,0,m} \hat{a}_{n',0,m'} B 
 (n,0,m;n',0,m') + {\rm h.c.} \right\} \biggr] \, , \label{H^0} \\ 
\hat{H}^{j'} & = \sum_{n,n',m,m'=0 }^{\infty} \biggl[
\hat{a}_{n,j',m}^{\dag} \hat{a}_{n',j',m'} A (n,j',m;n',j',m') \nn \\  
& \quad {} + \hat{a}_{n,-j',m}^{\dag} \hat{a}_{n',-j',m'}
 A(n,-j',m;n',-j',m') \nn \\  
& \quad {} + \frac{1}{2} \{ \hat{a}_{n,j',m} \hat{a}_{n',-j',m'} B
 (n,j',m;n',-j',m') + {\rm h.c.} \} \nn \\  
& \quad {} + \frac{1}{2} \{ \hat{a}_{n,-j',m} \hat{a}_{n',j',m'} B
 (n,-j',m;n',j',m') \nn \\ 
& \quad {} + {\rm h.c.} \} \biggr] \, , \label{Hprim}
\end{align}
with 
\begin{align}
 A(n,\ell,m;n',\ell',m') 
& =  \{ ( \varepsilon_{n,\ell,m} - \mu ) \delta_{n,n'} \delta_{m,m'} \nn
 \\ 
& \quad {} + 4 K(n,\ell,m;n',\ell',m') \} \delta_{\ell,\ell'} \, , \\ 
B(n,\ell,m;n',\ell',m') & = 2 K(n,\ell,m;n',\ell',m')
 \delta_{\ell,-\ell'} \, , 
\end{align}
where $K(n,\ell,m;n',\ell',m')$ is defined as 
\begin{align}
& K(n,\ell,m;n',\ell',m') \nn \\
& = \frac{g N_\mathrm{c}}{2} \int \! rdr \, dz \, f^2 (r,z) u_{n,\ell,m} (r,z)
 u_{n',\ell',m'} (r,z) \, . 
\label{defK}
\end{align}

\section{Rossignoli--Kowalski Method}\label{sec-RK}
In this section, we briefly review the RK method. In this method the 
``complex modes'' appear in the diagonalized Hamiltonian with
unusual operators which are neither bosonic nor fermionic ones
\cite{RK}.

We begin with the general Hamiltonian for the BEC of the quadratic form of 
creation and annihilation operators $\hat{a}_N^\dag$ and $\hat{a}_N$ as 
\begin{align}
\hat{H}_{\mathrm{quad}} & = \sum_{N, N'} \biggl\{ A_{NN'} \left(
 \hat{a}^{\dag}_N  
 \hat{a}_{N'} + \frac{1}{2}\delta_{N,N'} \right) \nn \\
& \quad {} + \frac{1}{2}(B_{NN'} \hat{a}^{\dag}_{N} \hat{a}^{\dag}_{N'}
 + {\rm h.c.}) \biggr\} \nn \\
& = \frac{1}{2} Z^{\dag} \mathcal{H} Z \, , 
\end{align}
where
\begin{align}
\mathcal{H} & = 
\begin{pmatrix}
A & B \\ 
B^\ast & A^t
\end{pmatrix}
 , \\ 
Z & = 
\begin{pmatrix}
\hat{a}  \\ 
\hat{a}^{\dag} 
\end{pmatrix}
.
\end{align}
Here, the matrices $A$ and $B$ are hermite and symmetric, respectively.
The symbols $\hat{a}$ and $\hat{a}^{\dag}$ represent 
the arrays of components $\hat{a}_N$
and $\hat{a}^{\dag}_N$:
\begin{align}
 \hat{a} & = 
\begin{pmatrix}
 \hat{a}_{N=1} \\
 \hat{a}_{N=2} \\
 \vdots
\end{pmatrix}
, \\
 \hat{a}^\dag & = 
\begin{pmatrix}
 \hat{a}_{N=1}^\dag \\
 \hat{a}_{N=2}^\dag \\
 \vdots
\end{pmatrix}
, 
\end{align}
where $N$ stands for the combination of the quantum numbers $n$, $\ell$
and $m$, and we treat $N$ as a single integer for simplicity. The
bosonic commutation relations of the original creation and annihilation
operators $\hat{a}_N$ and $\hat{a}_N ^{\dag}$ are written as  
\begin{equation}
Z Z ^{\dag} - ( Z^{\dag t} Z^t )^t = \mathcal{M} \, ,  
\end{equation}
where
\begin{equation}
\mathcal{M} = 
\begin{pmatrix}
1 & 0 \\ 
0 & - 1 
\end{pmatrix}
.
\end{equation}

Now, we consider a general linear canonical transformation of the 
$\hat{a}$ and $\hat{a}^{\dag}$
\begin{align}
Z & = \mathcal{W} Z' \, , \\   
Z' & = 
\begin{pmatrix}
 \hat{a}' \\
 \bar{a}'
\end{pmatrix}
 , 
\end{align}
where 
\begin{align}
 \hat{a}' & = 
\begin{pmatrix}
 \hat{a}'_{N=1} \\
 \hat{a}'_{N=2} \\
 \vdots
\end{pmatrix}
, \\
 \bar{a}' & = 
\begin{pmatrix}
 \bar{a}'_{N=1} \\
 \bar{a}'_{N=2} \\
 \vdots
\end{pmatrix}
.
\end{align}
We note that $\bar{a}'_N$ is not necessarily the hermitian conjugate of
$\hat{a}'_N$, although $\hat{a}'_N$ and $\bar{a}'_N$ are required to
satisfy the bosonic commutation relations as 
\begin{equation}
Z' \bar{Z}' - ( \bar{Z}'^t Z'^t )^t = \mathcal{M} \, ,
\end{equation}
where
\begin{align}
 \bar{Z}' & =  
\begin{pmatrix}
 \bar{a}' & \hat{a}'
\end{pmatrix}
 = Z'^t \mathcal{T} \, , \\
\mathcal{T} & = 
\begin{pmatrix}
0 & 1 \\ 
1 & 0  
\end{pmatrix}
. 
\end{align}
As $Z'^{\dag} = \bar{Z}' \bar{\mathcal{W}}$, the matrix $\mathcal{W}$
should satisfy 
\begin{equation}
\label{WM barW}
\mathcal{W M \bar{W}} = \mathcal{M} \, , 
\end{equation}
where 
\begin{equation}
\bar{\mathcal{W}} = \mathcal{T} \mathcal{W}^t \mathcal{T} \, .  
\end{equation}
It is important to notice that in the case where only the  real modes appear, 
the relation $\mathcal{\bar{W}} = \mathcal{W}^{\dag}$ holds, and then 
$\bar{a}'_N = \hat{a}'^{\dag}_{N}$. 

The quadratic Hamiltonian $\hat{H}_{\mathrm{quad}}$ is rewritten as 
\begin{align}
\hat{H}_{\mathrm{quad}} & = \frac{1}{2} \bar{Z}' \mathcal{H'} Z' \, , \\ 
\label{H'}
\mathcal{H'} & = \mathcal{\bar{W} H W} \, . 
\end{align}
Finding a representation in which  $\tilde{\mathcal{H'}}$ is diagonal
corresponds to solving an eigenvalue equation with the ``metric''
$\mathcal{M}$, i.e., 
\begin{equation}
\mathcal{H W} = \mathcal{MWMH'} \, .  
\end{equation}
We regard this equation as an eigenvalue equation for a
non-hermitian matrix $\tilde{\mathcal{H}}$: 
\begin{align}
\tilde{\mathcal{H}} \mathcal{W} & = \mathcal{W} \tilde{\mathcal{H'}} \,
 , \label{eigen_Htilde} \\ 
\tilde{\mathcal{H}} & = \mathcal{MH} = 
\begin{pmatrix}
A & B \\ 
- B^\ast & - A^t
\end{pmatrix}
, 
\end{align}
with the diagonal matrix $\tilde{\mathcal{H'}}$ whose diagonal
components are eigenvalues of $\tilde{\mathcal{H}}$, denoted by $\chi_N$. 
Then the quadratic Hamiltonian $\hat{H}_{\mathrm{quadratic}}$ can be
expressed as  
\begin{equation}
\hat{H}_{\mathrm{quad}} = \sum_N \chi_N \left( \bar{a}'_N \hat{a}'_N +
\frac{1}{2} \right) .  
\end{equation}

\section{Condition for the Existence of the Complex Modes} \label{sec:condition}

Applying the RK method \cite{RK} reviewed in the previous section
to our Hamiltonian in Eqs.~(\ref{H_sectors})--(\ref{defK}), we now
solve the eigenvalue equation~(\ref{eigen_Htilde}) under the circumstance
to obtain the mode-decoupled form of our Hamiltonian.  
But the equation is too difficult to solve 
analytically in general.

In this section, we develop the following two approximate schemes in
Eq.~(\ref{eigen_Htilde}). First, we make a two-mode approximation on the
Hamiltonian (\ref{H_sectors}). 
Second, we introduce the small coupling expansion.  Both of them are 
helpful for the analytical study in spite of the nonlinearity of
the basic equation  for the condensate (\ref{GP}).
And finally, we analytically derive the condition for the existence
of the complex modes. 

\subsection{Two-Mode Approximation}
In this subsection, we explain the two-mode approximation
which is used to obtain the analytic
expression of the condition for the existence of the complex modes. 

First, we consider the $j=0$ part of the free Hamiltonian
(\ref{H^0}). Within the two-mode approximation, $\hat{H}^{0}$  is written
in the matrix representation as 
\begin{equation}
\hat{H}^{0} = \frac{1}{2} Z_0 ^{\dag} \mathcal{H}_{0} Z_{0} \, , 
\end{equation}
 where 
\begin{align}
 Z_0 & = 
\begin{pmatrix}
 \hat{a}_1 \\ 
 \hat{a}_2 \\ 
 \hat{a}_1^{\dag} \\
 \hat{a}_2^{\dag}
\end{pmatrix}
, \\
Z_0^{\dag} & = 
\begin{pmatrix}
 \hat{a}_1^{\dag} & \hat{a}_2^{\dag} & \hat{a}_1 & \hat{a}_2 
\end{pmatrix}
, \\
\mathcal{H}_{0} & = 
\begin{pmatrix}
A_0 & B_0 \\ 
B_0 & A_0 
\end{pmatrix}
, \\ 
\label{A0}
A_0 & = 
\begin{pmatrix}
\varepsilon_{1} - \mu + 4 K(1,1) & 4K(1,2) \\ 
4K(2,1) & \varepsilon_{2} - \mu + 4 K(2,2)  
\end{pmatrix}
, \\ 
B_0 & = 
\begin{pmatrix}
2 K(1,1) & 2 K(1,2) \\ 
2 K(2,1) & 2 K(2,2) 
\end{pmatrix}
. 
\end{align}
Here, the mode 1 represents one labeled by the quantum numbers
$(n,\ell,m) = (n_1, 0, m_1)$ and the mode 2 does $(n,\ell,m) = (n_2, 0,
m_2)$.  
We also define the tilde matrix 
\begin{equation}
 \tilde{\mathcal{H}}_0 = 
\begin{pmatrix}
 A_0 & B_0 \\
 - B_0 & - A_0
\end{pmatrix} 
. \label{defcalH0}
\end{equation}

Next, we consider the $j \ne 0$ part of the free Hamiltonian
(\ref{Hprim}). In the matrix representation, $\hat{H}^{j'}$ under the
two-mode approximation is written as 
\begin{equation}
\hat{H}^{j'} = \frac{1}{2} Z_{j'}^{\dag} \mathcal{H}_{j'} Z_{j'} \, ,   
\end{equation}
where 
\begin{align}
 Z_{j'} & = 
\begin{pmatrix}
 \hat{a}_{1'} \\ 
 \hat{a}_{2'} \\ 
 \hat{a}_{1'}^{\dag} \\
 \hat{a}_{2'}^{\dag}
\end{pmatrix}
, \\
Z_{j'}^{\dag} & = 
\begin{pmatrix}
 \hat{a}_{1'}^{\dag} & \hat{a}_{2'}^{\dag} & \hat{a}_{1'} & \hat{a}_{2'} 
\end{pmatrix}
, \\
\mathcal{H}_{j'} & = 
\begin{pmatrix}
A_{j'} & B_{j'} \\ 
B_{j'} & A_{j'} 
\end{pmatrix}
, \\ 
\label{Aj} 
A_{j'} & = 
\begin{pmatrix}
\varepsilon_{1'} - \mu + 4 K(1',1') & 0 \\ 
0 & \varepsilon_{2'} - \mu + 4 K(2',2')  
\end{pmatrix}
, \\ 
B_{j'} & = 
\begin{pmatrix}
0 & 2 K(1',2') \\ 
2 K(2',1') & 0
\end{pmatrix}
. 
\end{align}
Here, the quantum numbers $(n,\ell,m)$ of the mode 1' and
the mode 2' are $(n_1', j', m_1' )$ and $(n_2', -j', m_2')$,
respectively. 
The tilde matrix is defined by
\begin{equation}
 \tilde{\mathcal{H}}_{j'} = 
\begin{pmatrix}
 A_{j'} & B_{j'} \\
 - B_{j'} & - A_{j'}
\end{pmatrix} 
. \label{defcalHj'}
\end{equation}

\subsection{Small coupling expansion}  
In this subsection, we define the dimensionless parameter $\lambda$ as
\begin{equation}
\lambda = d_{\mathrm{ho}} (g N_{\rm c}) , 
\end{equation}
and make an assumption of 
\begin{equation}
| \lambda | \ll 1 \label{smalllambda}, 
\end{equation}
where $d_{\mathrm{ho}} =(\alpha_{\bot}^2 \alpha_z)^{1/3} M$ is the
constant determined by the typical scales of the system,
and expand the function $f(r,z)$ (see Eq.~(\ref{GP2})) and 
chemical potential $\mu$ with respect to this small parameter $\lambda$. 
We refer 
to this expansion as the small coupling expansion. 
Here we perform this expansion up to the first order in $\lambda$. 
Note that this expansion is applicable not only for repulsive
interaction ($\lambda > 0$), but also for attractive one $(\lambda < 0)$.
Let us introduce the notations of quantities of the zeroth order in 
$\lambda$, namely, a real function $f_0 (r,z)$ and a chemical potential $\mu_0$,
which satisfy  
\begin{equation}
\label{f_0}
[ K_{\kappa} + V (r,z) ] f_0 (r,z) = \mu_0 f_0 (r,z) \, .  
\end{equation}
The function $f_0 (r,z)$ and the quantity $\mu_0$ correspond to the
eigenfunction $u_{0,0,0} (r,z)$ and the eigenvalue $\varepsilon_{0,0,0}$,
respectively, i.e.,
\begin{align}
f_0 (r,z) & =  \sqrt{2 \pi} \, u_{0,0,0} (r,z) \nn \\ 
& = \sqrt{2 \pi} \, C_{0,0,0} e^{- \frac{1}{2} \alpha_{\bot}^2
r^2} ( \alpha_{\bot} r )^{\kappa} e^{- \frac{1}{2} \alpha_z ^2 z^2} \, ,
 \\ 
\mu_0 & = \varepsilon_{0,0,0} = \omega_{\bot} ( \kappa + 1 ) +
 \frac{\omega_z}{2} \, . 
\end{align}
Note that the function $f_0(r,z)$ is normalized as 
\begin{equation}
\int r dr \, dz \, f_0^2(r,z)  =1. 
\end{equation}
Using the function $f_0 (r,z)$ we can expand the real function $f (r,z)$
up to the first order in $\lambda$: 
\begin{equation}
f(r,z) \simeq G \left\{ f_0(r,z) + \lambda f_1(r,z) \right\} \, ,
 \label{fexpansion} 
\end{equation}
where $f_1 (r,z)$ is a real function and normalized, 
\begin{equation}
 \int \! r dr \, dz \, f_1^2 (r,z) = 1 \, , 
\end{equation}
and the parameter $G$ is determined by the normalization condition
(\ref{normf}) as 
\begin{equation}
G = (1+ 2 \lambda p_1)^{-\frac{1}{2}} \, ,
\end{equation}
where the coefficient $p_1$ is defined as 
\begin{equation}
p_1 = \int \! rdr \, dz \, f_0 (r,z) f_1 (r,z) \, ,  
\end{equation}
which leads to
\begin{equation}
\label{f}
f (r,z) \simeq f_0 (r,z) + \lambda\{ f_1 (r,z) - p_1 f_0 (r,z) \} \, . 
\end{equation}
We also need to expand the chemical potential $\mu$ up to the first order
in $\lambda$: 
\begin{equation}
\label{mu}
\mu \simeq \mu_0 + \lambda \mu_1 \, .  
\end{equation}
We determine $\mu_1$ by substituting Eq.~(\ref{f}) and Eq.~(\ref{mu})
into the GP equation (\ref{GP2}), 
\begin{equation}
 \{ K_{\kappa} + V (r,z) - \mu_0 \} f_1 (r,z) - \mu_1 f_0 (r,z)
 +  \frac{1}{2 \pi d_{\mathrm{ho}}}f_0^3 (r,z) = 0 \, .   \label{f_1-eq.} 
\end{equation}
Here we expand the function $f_1 (r,z)$ with the eigenfunctions
$u_{n,0,m} (r,z)$, 
\begin{equation}
\label{f_1}
f_1 (r,z) = \sum_{n,m=0}^{\infty} s_{n,m} u_{n,0,m} (r,z) \, ,
\end{equation}
where the expansion coefficients $s_{n,m}$ are written as 
\begin{equation}
s_{n,m} = 2 \pi \int \! r dr \, dz \, u_{n,0,m} (r,z) f_1 (r,z) \, .  
\end{equation}
It is notable to see that
\begin{equation}
s_{0,0} = \sqrt{2 \pi} \, p_1 \, . 
\end{equation}
Substitute Eq.~(\ref{f_1}) to Eq.~(\ref{f_1-eq.}), one obtains 
\begin{align}
& \sum_{n,m=0}^{\infty} (\varepsilon_{n,0,m} - \mu_0 )s_{n,m} u_{n,0,m}
 (r,z) - \mu_1 f_0 (r,z) \nn  \\ 
\label{mu1-1}
& {} + \frac{1}{2 \pi d_{\mathrm{ho}}} f_0^3 (r,z) = 0 \, .  
\end{align}
Multiplying the function $f_0 (r,z)$ to Eq.~(\ref{mu1-1}) and
integrating over the space coordinates, we obtain 
\begin{equation}
\mu_1 = \frac{1}{2 \pi d_{\mathrm{ho}}} \int \! rdr \, dz \, f_0^4 (r,z) 
= \frac{2 \pi}{d_{\mathrm{ho}}} C \frac{(2 \kappa )!}{(\kappa !)^2} 
\end{equation}
where $C$ is defined as 
\begin{equation}
C = \sqrt{\frac{1}{2 \pi^5}} \cdot \frac{\alpha_{\bot}^2 \alpha_z}{2^{2
 \kappa + 2}} \, .   
\end{equation}
Thus $f_1(r,z)$ and $\mu_1$ are determined.

\subsection{The condition for the existence of the complex modes}
In this subsection, we calculate the eigenvalues of the matrices
$\tilde{{\cal H}}_0$ in Eq.~(\ref{defcalH0}) and $\tilde{\mathcal{H}}_{j'}$ 
in Eq.~(\ref{defcalHj'}) under the small
coupling expansion.
To simplify the expression, we parameterize the elements of the matrices 
$A_{j}$ and $B_{j}$ $(j = 0, j')$ as 
\begin{align}
A_{j,11} & = a_j + r_{j,a} \, , \\
A_{j,22} & = a_j - r_{j,a} \, , \\
A_{j,12} & = A_{j,21} = \Delta_{j,a} \, , \\
B_{j,11} & = b_j + r_{j,b} \, , \\
B_{j,22} & = b_j - r_{j,b} \, , \\
B_{j,12} & = B_{j,21} = \Delta_{j,b} \, . 
\end{align}
The eigenvalues $\chi_j$ of the matrix $\mathcal{\tilde{H}}_j$ are 
characterized by the equation
\begin{equation}
 \det [ \mathcal{\tilde{H}}_j - \chi_j ] = 0 \, . 
\end{equation}
Then we obtain 
\begin{align}
\chi^2_j & =  (a_j^2 + r_{j,a}^2 + \Delta_{j,a}^2 )
- (b_j^2 + r_{j,b}^2 + \Delta_{j,b}^2 ) \pm 2 \sqrt{I} \, , \\ 
I & = ( a_j r_{j,a} - b_j r_{j,b} )^2 
+ ( a_j \Delta_{j,a} - b_j \Delta_{j,b} )^2 \nn \\ 
& \quad {} - ( r_{j,a} \Delta_{j,b} - r_{j,b} \Delta_{j,a} )^2 \, . 
\end{align}

We discuss the two cases, $j=0$ and $j \ne 0$, separately.

\subsubsection{$j=0$ case}
In the case of $j=0$, the elements of the matrix $A_0$ are written as
\begin{align}
a_0 & = \frac{1}{2} \{ 2\omega_{\bot} ( n_1 + n_2 ) + \omega_z ( m_1 +
 m_2 ) \} + O (\lambda) \label{a01} \\ 
& = \bar{a}_0 + O (\lambda) \, , \\
r_{0,a} & = \frac{1}{2} \{ 2 \omega_{\bot} ( n_1 - n_2 ) + \omega_z (
 m_1 - m_2 ) \} + O (\lambda) \\
& = \bar{r}_{0,a} + O (\lambda) \, . 
\end{align}
It is easy to see that if $| \lambda | \ll 1$, then 
\begin{equation}
\chi^2_0 = (\bar{a}_0 \pm \bar{r}_{0,a})^2 + O (\lambda) > 0 \, .  
\end{equation}
>From this relation, we conclude that no complex mode appears
for $j=0$ and sufficiently small $\lambda$.

\subsubsection{$j\neq0$ case}
In the case of $j \ne 0$, it is straightforward to see 
\begin{equation}
\chi^2_{j'} = \left( r_{j',a} \pm \sqrt{a_{j'}^2 - \Delta_{j',b}^2 }
\right)^2 
\end{equation}
for $j' \ge 1$. 
So the sign of $a_{j'}^2 - \Delta_{j',b} ^2$ determines whether the
eigenvalues are real or not. 
Note that it has the term of $\varepsilon_{1'} +
\varepsilon_{2'}$, and when we search complex eigenvalues, it
is natural to choose the mode 1' as $(0,j',0)$ and the 
mode 2' as $(0,-j',0)$,
because then $a^2_{j'}$ reaches its minimum.

In the case of $j' > \kappa $, $\varepsilon_{1'}$ and $\varepsilon_{2'}$
are written as  
\begin{align}
\varepsilon_{1'} & = \omega_{\bot} ( j' + \kappa + 1 ) +
 \frac{\omega_z}{2}, \\ 
\varepsilon_{2'} & = \omega_{\bot} ( j' - \kappa + 1 ) +
 \frac{\omega_z}{2} \, .
\end{align}
Since the matrix element $a_{j'}$ is written as 
\begin{equation}
a_{j'} = \omega_{\bot} ( j' - \kappa ) - \lambda \mu_1 + 2 K(1',1') +
 2K(2',2') \, , \label{aj1}
\end{equation}
the term $a_{j'}^2 - \Delta_{j',b}^2$ is evaluated as 
\begin{equation}
a_{j'}^2 - \Delta_{j',b} ^2 = \omega_{\bot} ^2 ( j' - \kappa )^2 +
 O(\lambda) > 0 \, .  
\end{equation}
So we can see that in the case of $j' > \kappa$ no complex mode
exists.

In the case of $0 < j' \le \kappa$, $\varepsilon_{1'}$ and
$\varepsilon_{2'}$ are
\begin{align}
\varepsilon_{1'} & = \omega_{\bot} ( j' + \kappa + 1 ) +
 \frac{\omega_z}{2} \, , \\
\varepsilon_{2'} & = \omega_{\bot} ( - j' + \kappa + 1 ) +
 \frac{\omega_z}{2} \, .  
\end{align}
Then the matrix element $a_{j'}$ becomes
\begin{equation}
a_{j'} = - \lambda \mu_1 + 2 K(1',1') + 2 K(2',2') 
\end{equation}
where $K(1',1')$ and $K(2',2')$ are written as 
\begin{align}
K (1',1') & =\frac{2 \pi}{d_{\mathrm{ho}}} \lambda
 \frac{C}{2} \frac{ ( 2 \kappa + j' )! }{2^{j'}
 (\kappa + j')! \kappa !} \, , \\ 
K (2',2') & =\frac{2 \pi}{d_{\mathrm{ho}}}
 \lambda  \frac{C}{2} \frac{ ( 2 \kappa - j' )! }{2 ^{-j'}
 (\kappa -j')! \kappa !} \, . 
\end{align}
The parameter $\Delta_{j',b}$ is given as 
\begin{equation}
\Delta_{j',b} =\frac{2 \pi}{d_{\mathrm{ho}}} \lambda C
\frac{ ( 2 \kappa )! }{\kappa ! \sqrt{(\kappa + j')! (\kappa - j')! } }
\, . 
\end{equation}
It is important to see that $a_{j'}$ and $\Delta_{j',b}$ have no term
of the zeroth order in the parameter $\lambda$ (cf. Eqs.~(\ref{a01}) and
(\ref{aj1})). 
Thus we must evaluate all terms
in $a_{j'}^2 - \Delta_{j',b}^2$ up to the second order
in $\lambda$.
Then, $a_{j'}^2 - \Delta_{j',b}^2$ is evaluated as
\begin{align}
 a_{j'}^2 - \Delta_{j',b}^2 & = \lambda^2 \left\{ \mu_1 - \frac{2
 \pi}{d_{\mathrm{ho}}} C (S + T) \right\} \nn \\
 & \quad {} \times \left\{ \mu_1 - \frac{2 \pi}{d_{\mathrm{ho}}} C (S -
 T) \right\} \, , 
\end{align}
where $S$ and $T$ are defined as 
\begin{align}
S & = \frac{( 2 \kappa + j' )!}{2^{j'} (\kappa + j')! \kappa !} 
+ \frac{ ( 2 \kappa - j' )! }{2 ^{-j'} (\kappa -j')! \kappa !} \, , \\ 
T & = \frac{ ( 2 \kappa )! }{\kappa ! \sqrt{(\kappa + j')! (\kappa -
 j')! } } \, .  
\end{align}
Therefore, the condition for the existence of the complex modes,
 $a_{j'}^2 - \Delta_{j',b} ^2 < 0$, now amounts to
\begin{equation}
\label{condition}
S - T < \frac{(2 \kappa)!}{(\kappa !)^2} < S + T \, . 
\end{equation}

\begin{table}
\begin{tabular}[t]{|c||p{4mm}|p{4mm}|p{4mm}|p{4mm}|p{4mm}|p{4mm}|}
 \hline
 $\kappa \backslash j$ & \multicolumn{1}{|c|}{1} &
 \multicolumn{1}{|c|}{2} & \multicolumn{1}{|c|}{3} &
 \multicolumn{1}{|c|}{4} & \multicolumn{1}{|c|}{5} & $\cdots$ \\ 
\hline \hline
 1 & & & & & & \\
\hline
 2 & & \multicolumn{1}{|c|}{$\times$} & & & & \\ 
\hline
 3 &  & \multicolumn{1}{|c|}{$\times$} & \multicolumn{1}{|c|}{$\times$}
 & & & \\
\hline
 4 & & \multicolumn{1}{|c|}{$\times$} & \multicolumn{1}{|c|}{$\times$} &
 & & \\
\hline
 5 & & \multicolumn{1}{|c|}{$\times$} & \multicolumn{1}{|c|}{$\times$} &
 \multicolumn{1}{|c|}{$\times$} & & \\
\hline
 6 & & \multicolumn{1}{|c|}{$\times$} & \multicolumn{1}{|c|}{$\times$} &
 \multicolumn{1}{|c|}{$\times$} & & \\
\hline
 $\vdots$ &  &  & & & & \\
\hline
\end{tabular}
\caption{The existence of the complex modes. 
The condition (\ref{condition}) are examined for various 
values of $(\kappa, j)$.
The marks are the positions where the complex modes arise. 
This result is valid for the repulsive interaction ($\lambda > 0$), 
as well as the attractive one ($\lambda < 0$), 
under the situation of $|\lambda| \ll 1$.
}
\label{cetable}
\end{table}

We apply the condition~(\ref{condition}) to each winding number and
find the appearance of the complex modes (Table \ref{cetable}). In the
case of the winding number $\kappa =1$, no complex mode exists.
On the other hand, in the case of the winding number $\kappa \ge 2$ the
complex modes always exist. Particularly, we find that in the case of
the winding number $\kappa =2$ the complex modes appear as a pair, 
while in case of $\kappa =3$ the complex modes arise as two pairs. So
our results in the case of the winding number $\kappa =2$ and $3$ are
consistent with those by Pu {\it et al.}~\cite{Pu} and 
M\"ott\"onen {\it et al.}~\cite{Mettenen}. 
In the case $\kappa =4$, our result is consistent with that by 
Kawaguchi and Ohmi \cite{Kawaguchi}. 
Remark that Pu  {\it et al.}~\cite{Pu} performs their analysis
also for the attractive interaction ($\lambda < 0$), 
while the other authors do only for the repulsive interaction ($\lambda > 0$)
\cite{Mettenen, Kawaguchi}.
Our result is also consistent with that 
by Pu {\it et al.}~\cite{Pu} for $\lambda < 0$.

The condition~(\ref{condition}) can be applicable only in the small
coupling constant region. So the existence of new complex modes 
which may arise in the large coupling
constant region, such as one can find them in Fig.~2 in
Ref.~\cite{Kawaguchi}, is outside the scope of this paper. 

\subsection{Proof of the existence of the complex modes for $\kappa \ge 2$}

One of the most important consequence of the analytic result (\ref{condition})
is the following statement: {\it The $j' = 2$ modes are always complex
ones for $\ka \ge 2$.}
The proof is given below. 

For $j' = 2$, the inequality (\ref{condition}) is reduced to the 
following coupled inequalities
\begin{equation}
 1<
\frac{(2 \ka + 2) (2\ka +1)}{4(\ka + 2)(\ka + 1) } + \frac{4 \ka (\ka
-1)}{ (2 \ka)(2\ka -1)} + \sqrt{\frac{\ka (\ka -1)}{(\ka + 2) (\ka +
1)}} \, , \label{ineq2}
\end{equation}
\begin{equation}
 \frac{(2 \ka + 2) (2\ka +1)}{4(\ka + 2)(\ka + 1) } + \frac{4 \ka (\ka
 -1)}{ (2 \ka)(2\ka -1)} - \sqrt{\frac{\ka (\ka -1)}{(\ka + 2) (\ka +
 1)}} < 1 \, . \label{ineq1}
\end{equation}

First, let us prove the inequality (\ref{ineq2}). 
Remark that the inequality
\begin{equation}
\frac{(2 \ka + 2) (2\ka +1)}{4(\ka + 2)(\ka + 1) } + \frac{4 \ka (\ka
 -1)}{ (2 \ka)(2\ka -1)} - 1 > 0   
\end{equation}
is reduced to 
\begin{equation}
\ka (\ka +1) (4 \ka^2 -2 \ka -5) > 0 \, .  
\end{equation} 
The largest root for the left-hand side is $\ka = \kappa_1 \simeq
1.396$. 
Combining this result with the fact
$\sqrt{\{\ka (\ka -1)\} / \{(\ka + 2) (\ka + 1)\}} > 0$ for $\ka >
\kappa_1$, we can find the inequality (\ref{ineq2}) is satisfied for
$\kappa > \kappa_1$. 

Next, let us prove the inequality (\ref{ineq1}). 
The inequality (\ref{ineq1}) is equivalent to
\begin{equation}
\ka^2 (\ka+1) ( 8\ka^3 + 52 \ka^2 -53 \ka -25 ) > 0 \, . 
\end{equation}
The largest root for the left-hand side is $\ka = \kappa_2 \simeq 1.199
$. 
>From the same discussion on the inequality (\ref{ineq2}), we find that the
inequality (\ref{ineq1}) is satisfied by $\kappa > \kappa_2$. 

>From our discussion above, the roots $\kappa_1$ and $\kappa_2$ satisfy
the inequality $1 < \kappa_1, \kappa_2 < 2$. 
This means that if $\kappa \ge 2$, $\kappa$ satisfies the inequalities
(\ref{ineq2}) and (\ref{ineq1}) simultaneously, then the statement is
proven. 

\section{Summary}\label{summary}
We derived the analytic expression of the condition for the existence of
the complex modes when the condensate has a highly quantized vortex,
using the method developed by Rossignoli and Kowalski \cite{RK} for the
small coupling constant, under the two-mode approximation. 
Our results agree with those by Pu {\it et al.}~($\kappa =2,3$)
\cite{Pu}, M\"ott\"onen {\it et al.}~($\kappa = 2$) \cite{Mettenen} and  
Kawaguchi and Ohmi ($\kappa = 4$) \cite{Kawaguchi}. 
It is emphasized that the formula derived here is applicable for an
arbitrary winding number, and one can check whether complex modes exist
or not for an arbitrary highly quantized vortex in the condensate. 
We have also proven that in the case of $\kappa \ge 2$, the complex
modes always exist. 
This means that the arbitrarily highly quantized vortex may be dynamically
unstable. 
Moreover, it is shown that imaginary parts of the complex eigenvalues
rise linearly against the coupling constant. 

The validity of the two-mode approximation is certainly not clear.
But the numerical calculation for the case of $\kappa = 2$ and $\lambda
> 0$ revealed that the behavior of the imaginary part in the small
coupling constant region is not significantly affected by the number of
the modes included to the calculation~\cite{Pu}.  
This fact suggests that the two-mode approximation does not modify
behaviors of complex modes in the small coupling constant region
essentially and that the analysis in this paper is reliable.

Finally, we comment that the physical interpretation of the complex
modes is very difficult and is not settled yet. Pu {\it et al.}~employs  
the classical interpretation on the complex modes, with the idea that
the complex modes cause the dynamical instability, that is, the decay of
the initial configuration of the condensate \cite{Pu}. The argument does
not touch the aspect of quantum theory. We recently proposed a quantum
field theoretical treatment associated with the complex modes and sought
a new interpretation \cite{Mine}.

\begin{acknowledgments}
The authors would like to thank Professor I.~Ohba and Professor
 H.~Nakazato for helpful comments and encouragements. 
M.M. and T.S. are supported partially by the Grant-in-Aid for The 21st
Century COE Program (Physics of Self-organization Systems) at Waseda
University.
This work is partly supported by a Grant-in-Aid for Scientific Research
(C) (No.~17540364) from the Japan Society for the Promotion of Science,
for Young Scientists (B) (No.~17740258) and for
Priority Area Research (B) (No.~13135221) both from the Ministry of
Education, Culture, Sports, Science and Technology, Japan.
\end{acknowledgments}



\begin{thebibliography}{99}
\bibitem{Boulder}
M. H.~Anderson, J. R.~Ensher, M. R.~Matthews, C. E.~Wieman, and
E. A.~Cornell, Science {\bf 269}, 198 (1995).

\bibitem{MIT}
K. B.~Davis, M. -O.~Mewes, M. R.~Andrews, N. J.~van Druten, D. S.~Durfee,
D. M.~Kurn, and W.~Ketterle, Phys. Rev. Lett. {\bf 75}, 3969 (1995).

\bibitem{Li}
C. C.~Bradley, C. A.~Sackett, J. J.~Tollett, and R. G.~Hulet,
Phys. Rev. Lett. {\bf 75}, 1687 (1995).

\bibitem{Matthews}
M.~R.~Matthews, B.~P.~Anderson, P.~C.~Haljan, D.~S.~Hall,
C.~E.~Wieman, and E.~A.~Cornell, Phys. Rev. Lett. {\bf 83}, 2498 (1999).

\bibitem{Madison}
K.~W.~Madison, F.~Chevy, W.~Wohlleben, and J.~Dalibard,
Phys. Rev. Lett. {\bf 84}, 806 (2000).

\bibitem{Abo-Shaeer}
J.~R.~Abo-Shaeer, C.~Raman, J.~M.~Vogels, and W.~Ketterle,
Science {\bf 292}, 476 (2001)

\bibitem{Leanhardt}
A.~E.~Leanhardt, A.~G\"orlitz, A.~P.~Chikkatur, D.~Kielpinski,
Y.~Shin, D.~E.~Pritchard, and W.~Ketterle,
Phys. Rev. Lett. {\bf 89}, 190403 (2002).

\bibitem{Nakahara}
M.~Nakahara, T.~Isoshima, K.~Machida, S.-I.~Ogawa, and T.~Ohmi, Physica
B {\bf 284}, 17 (2000); T.~Isoshima, M.~Nakahara, T.~Ohmi, and
K.~Machida, Phys. Rev. A {\bf 61}, 063610 (2003); S.-I.~Ogawa, M.~M{\"
o}tt{\" o}nen, M.~Nakahara, T.~Ohmi, and H.~Shimada, {\it ibid}. {\bf
66}, 013617 (2002); M.~M{\" o}tt{\" o}nen, N.~Matsumoto, M.~Nakahara,
and T.~Ohmi, J. Phys:Condens. Matter {\bf 14}, 29 (2002). 

\bibitem{Shin}
Y.~Shin, M.~Saba, M.~Vengalattore, T.~A.~Pasquini, C.~Sanner,
A.~E.~Leanhardt, M.~Prentiss, D.~E.~Pritchard, and W.~Ketterle,
Phys. Rev. Lett. {\bf 93}, 160406 (2004).

\bibitem{Pu}
H.~Pu, C.~K.~Law, J.~H.~Eberly, and N.~P.~Bigelow, Phys. Rev. A
{\bf 59}, 1533 (1999).

\bibitem{Mettenen}
M.~M{\" o}tt{\" o}nen, T.~Mizushima, T.~Isoshima, M.~M.~Salomaa, and
K.~Machida, Phys. Rev. A {\bf 68}, 023611 (2003).

\bibitem{Kawaguchi}
Y.~Kawaguchi and T.~Ohmi, Phys. Rev. A. {\bf 70}, 043610 (2004).

\bibitem{Bogoliubov}
N.~N.~Bogoliubov, J. Phys. (Moscow) {\bf 11}, 23 (1947). 

\bibitem{deGennes}
P.~G.~de Gennes, {\it Superconductivity of Metals and Alloys}
(Benjamin, New York, 1966).

\bibitem{Fetter}
A.~L.~Fetter, Ann. of Phys. {\bf 70}, 67 (1972).


\bibitem{Gawryluk}
K.~Kawryluk, M.~Brewczyk, and K.~Rz\c{a}\.zewski, J. Phys. B:
At. Mol. Opt. Phys. {\bf 39}, L225 (2006). 

\bibitem{Huhtameki}
J.~A.~M.~Huhtam\"aki, M.~M\"ott\"onen, T.~Isoshima, V.~Pietil\"a, and
S.~M.~M.~Virtanen, Phys. Rev. Lett. {\bf 97}, 110406 (2006). 

\bibitem{RK}
R.~Rossignoli and A.~M.~Kowalski, Phys. Rev. A {\bf 72}, 032101 (2005). 

\bibitem{Mine}
M.~Mine, M.~Okumura, T.~Sunaga, and Y.~Yamanaka, cond-mat/0609052. 

\bibitem{LH}
We use the following convensions: 
$L_n^k(x) = e^x \frac{x^{-k}}{n!} \frac{d^n}{dx^n}(e^{-x}x^{n+k})$, 
$H_m(x)=(-1)^m e^{x^2} \frac{d^m}{dx^m} e^{-x^2}$.

\end{thebibliography}
\end{document}